\documentclass[12pt]{iopart}
\usepackage{iopams}
\usepackage[T1]{fontenc}
\usepackage[latin9]{inputenc}
\usepackage{graphicx}
\usepackage{booktabs}
\usepackage{array}
\usepackage{xspace}

\usepackage{array}
\usepackage{color}

\usepackage{sidecap}

\newcommand{\nn}{\nonumber}

\usepackage[nolist]{acronym}

\makeatletter

\newcommand{\eqref}[1]{(\ref{#1})}

\usepackage{graphicx}
\usepackage{booktabs}

\newcommand{\ie}{i.\,e.\@\xspace}

\newcommand{\Arg}{\mathrm{Arg}}

\newcommand{\shalf}{{\textstyle \frac{1}{2}}}

\makeatother

\begin{document}

\title[Chaos synchronization in networks of delay-coupled lasers]
{Chaos synchronization in networks of delay-coupled lasers: Role of the coupling phases }

\author{Valentin Flunkert$^{1,2}$}


\author{Eckehard Sch{\"o}ll$^{1}$}

\address{$^{1}$Institut f{\"u}r Theoretische Physik, Technische
  Universit{\"a}t Berlin,\\
  Hardenbergstra\ss{}e 36, 10623 Berlin, Germany}

\address{$^{2}$Instituto de Fisica Interdisciplinar y Sistemas
  Complejos,\\
  IFISC (UIB-CSIC), Campus Universitat de les Illes Balears, E-07122
  Palma de Mallorca, Spain}

\ead{flunkert@itp.tu-berlin.de}

\begin{abstract}
  We derive rigorous conditions for the synchronization of
  all-optically coupled lasers. In particular, we elucidate the
  role of the optical coupling phases for synchronizability
  by systematically discussing all possible network motifs containing
  two lasers with delayed coupling and feedback.
  Hereby we explain previous experimental findings.
  Further, we study larger networks
  and elaborate optimal conditions for chaos synchronization.
  We show that the relative phases between lasers can be used to
  optimize the effective coupling matrix.
\end{abstract}



\maketitle

\section{Introduction}

Coupled nonlinear system may exhibit a remarkable phenomenon called
chaos synchronization \cite{PIK01,BOC02}. The individual systems
synchronize to a common chaotic trajectory.  This phenomenon has
received much attention, due to its potential applications in secure
communication \cite{CUO93,KAN08a}.  For technological applications
semi-conductor lasers are promising systems to implement secure
communication schemes using chaos synchronization, because they
exhibit fast dynamics, they are cheap and one could utilize the
existing telecommunication infrastructure for these lasers
\cite{ARG05}.

However, due to the fast dynamics of the lasers, propagation distances
of already a few meters introduce non-negligible delay times in the
coupling. The synchronization of delay-coupled systems in general
\cite{DHA04,CHO09,KIN09,KAN11} and in particular delay-coupled lasers
\cite{WUE05a,CAR06,ERZ06a,FIS06,DHU08,FLU09,ZAM10,HIC11,AVI12} has
thus been a focus of research in nonlinear dynamics during the last
decades.

When the lasers are coupled all-optically, not only delay effects are
important, but also the optical coupling phases of the coherently
coupled electric fields play an important role.  Coherent coupling may
result in constructive or destructive interference of incoming
signals. When the lasers are synchronized, this interference can occur
even if the coupling distance is much larger than the coherence length
of the beams \cite{AVI08}.

It has been demonstrated experimentally and numerically that by tuning
coupling phases one can adjust the level of synchronization ranging
from perfect synchronization to almost no correlation \cite{PEI02}.
However, the impact of these coupling phases and the resulting
interference conditions have so far not been thoroughly investigated.

In this work we derive and discuss synchronization conditions for
all-optically coupled lasers. These conditions ensure the existence
and stability of a completely synchronized solution, i.e., perfect
synchronization of identical lasers. We confirm the analytical
results by numerical simulations.  For our stability analysis we
employ recent results concerning the \ac{MSF} \cite{PEC98} in the
limit of large delay times \cite{FLU10b,HEI11}.

The paper is organized as follows. In Sec.~\ref{sec:review} we briefly
review previous results about synchronization in delay-coupled
networks in a general context. These general results are then applied
to all-optically coupled lasers. In particular, we discuss the case of
two all-optically coupled lasers in detail in
Sec.~\ref{sec:twolasers}. Here, we consider different coupling schemes
(unidirectional and bidirectional, open- and closed-loop) and derive
the corresponding synchronization conditions.  We then discuss
synchronization in larger laser networks in Sec.~\ref{sec:networks}
focusing on an experimentally feasible setup. Finally, we conclude
with Sec.~\ref{sec:conclusions}.

\section{Synchronization of delay-coupled systems}
\label{sec:review}

Consider a delay-coupled network of $N$ identical units
\begin{equation}
  \dot x_k(t) = f(x_k(t)) + \sum_{j=1}^N G_{kj} \, h(x_j(t-\tau)) \;, \label{eq:network}
\end{equation}
where $x_k\in\mathbb{R}^n$ is the state vector of the $k$-th node
($k=1,\dots,N$). Here, $f$ is a function describing the local dynamics
of an isolated node, $G_{kj}$ is a coupling matrix that determines the
coupling topology and the strength of each link in the network, $h$ is
a coupling function, and $\tau$ is the delay time in the connection,
which is assumed to be equal for all links.  A necessary condition for
perfect synchronization is that the matrix $G$ has a constant row sum
\begin{equation}
  {\bf C1:}\qquad \sigma = \sum_{j=1}^N G_{kj} \;, \label{eq:C1}
\end{equation}
independent of $k$.  This condition ensures that an invariant
synchronization manifold exists.  Note that in this case the row sum
$\sigma$ is an eigenvalue of the matrix $G$ corresponding to the
eigenvector $(1,1,\dots,1)$ of synchronized dynamics.  For a given
constant row sum $\sigma$ the dynamics in the synchronization manifold
of Eq.~\eqref{eq:network} is
\begin{equation}
  \dot{\bar{x}}(t) = f[\bar{x}(t)] + \sigma h[\bar{x}(t-\tau)] \;. \label{eq:syncdyn}
\end{equation}
where $\bar{x}(t)=x_1(t)=x_1(t)= \dots=x_N(t)$ denotes the synchronized solution.

Note that condition C1 addresses mathematically perfect
synchronization. For realistic systems it is important to investigate
what happens to the synchronization manifold under small perturbations
of the perfect setup, such as parameter mismatch of the individual
elements or coupling strength mismatch leading to a slightly broken
condition C1.  On one hand, there has been some recent progress in
addressing parameter mismatch within the \ac{MSF} framework
\cite{SUN09a,SOR11b} and it might be interesting to apply these new
methods to the laser system. On the other hand, slight perturbations
of the perfect setup are also related to the question whether the
synchronization manifold is normally hyperbolic
\cite{KOC00a}. Although these questions are important for realistic
systems, we will focus our analytic investigations on the case of
perfect synchronization to highlight our main message: the importance
of the coupling phases.

As we will discuss below, the necessary condition C1 becomes quite
involved for the case of optically coupled lasers, because here the
matrix $G$ is complex valued due to optical phase factors in the
coupling and, additionally, the lasers can have relative phase shifts,
which effectively allow the system to adjust the coupling matrix to
some extent.

If condition C1 is satisfied and a synchronization manifold exists,
the stability problem of the synchronized solution can be approached
using the \ac{MSF} \cite{PEC98}.
The \ac{MSF} depends on a complex
parameter $re^{i\psi}$ and is defined as the largest Lyapunov exponent
$\lambda(re^{i\psi})$ arising from the variational equation
\begin{equation}
  \dot{\xi}(t) = Df[\bar{x}(t)]\, \xi(t) + r e^{i\psi} Dh[\bar{x}(t-\tau)]\,\xi(t-\tau)\;,
\end{equation}
where $\bar{x}(t)$ is the synchronized trajectory of the system
determined by Eq.~\eqref{eq:syncdyn} and $Df$ and $Dh$ are Jacobians.

The synchronized state is stable for a given coupling topology if the
\ac{MSF} is negative at all transversal eigenvalues $\gamma_k$ of the
coupling matrix ($\lambda(\gamma_k)<0$).  Here, {\em transversal
  eigenvalue}\/ refers to all eigenvalues except for the eigenvalue
$\sigma$ associated to perturbations within the synchronization
manifold with corresponding eigenvector $(1,\,1,\,\dots,\,1)$.

The dynamical time scale of a semiconductor laser is given by the
relaxation oscillation period $T_{RO}$. Typical values of $T_{RO}$ for
semiconductor lasers are between $10^{-10}\textrm{s}$ and
$10^{-9}\textrm{s}$. This time scale corresponds to about
$3\textrm{cm}$ -- $30\textrm{cm}$ of optical path-length in air.  The
coupling distances between the lasers is usually of the order of
meters or even much larger \cite{ARG05}.  In this case the coupling
delay time is much larger than the intrinsic time scale $\tau\gg
T_{RO}$ of the lasers. The case of large delay times is an import
limit for delayed systems \cite{FAR82,GIA96,MEN98b,WOL06,YAN06,YAN09}.

We recently showed \cite{FLU10b} that in any network the stability
problem for the synchronized state is drastically simplified in this
limit of large delay times:
\begin{itemize}
\item The \ac{MSF} is rotationally symmetric around the origin in the
  complex plane, i.e., $\lambda(r\,e^{i\psi})$ is independent of $\psi$.
\item If $\lambda(0)>0$, then $\lambda(re^{i\psi})=\lambda(0)$ for all $r$
  (and $\psi$).
\item If $\lambda(0)<0$, then the \ac{MSF} is monotonically growing with respect to the
  parameter $r$ and there is a critical radius $r_0$, where it changes
  sign ($\lambda(r_0)=0$).
\end{itemize}
Recently, this structure of the \ac{MSF} has been
been confirmed experimentally \cite{ILL11} and even utilized to
predict synchronizability of a network from a simpler motif
\cite{ILL11,FLU11a}.

Note that the two cases $\lambda(0) > 0$ and $\lambda(0) < 0$ have recently
been discussed in more detail \cite{HEI11}. The case $\lambda(0) < 0$
is called {\em weak chaos}, since in this case the maximum Lyapunov
exponent scales as $\sim 1/\tau$ for $\tau\to\infty$. The case
$\lambda(0)>0$, on the other hand, is called {\em strong chaos}\/
since the maximum Lyapunov exponent scales as $\mathcal{O}(1)$ for
$\tau\to\infty$. It has been shown that for large delay networks can
only exhibit chaos synchronization in the regime of weak chaos
\cite{FLU10b,HEI11}.

The structure of the \ac{MSF} allows us to draw general conclusions
about the synchronizability of a given network topology. In
particular, chaos synchronization can only be stable if $\sigma$ is
the eigenvalue of $G$ with largest magnitude, i.e., the transversal
eigenvalues $\gamma_1,\, \gamma_2,\, \dots \gamma_{N-1}$ have smaller
magnitude
\begin{equation}
  {\bf C2:}\qquad |\gamma_n|<|\sigma| \;. \label{eq:C2}
\end{equation}
This condition (C2) is necessary for chaos synchronization
($|\sigma|>r_0$) and it is
sufficient for synchronization on a periodic orbit ($|\sigma|<r_0$) \cite{FLU10b}. In
fact the smaller the magnitude of the transversal eigenvalues, the
easier it is to synchronize the system.

For given synchronized dynamics (Eq.~\eqref{eq:syncdyn}), i.e.,
given system parameters and row sum $\sigma$, one can calculate
numerically the critical radius $r_0$, which then provides a necessary
and sufficient condition (C3) for synchronization (provided C1 is
fulfilled)
\begin{equation}
  {\bf C3:}\qquad |\gamma_n| < r_0\;. \label{eq:C3}
\end{equation}

While condition C3 is necessary and sufficient, its disadvantage
is that one needs to calculate $r_0$ explicitly for the particular
synchronized dynamics.  In contrast,
condition C2 is only necessary for chaotic synchronization, however,
its strengths lies in the fact that it depends solely on the coupling
topology, i.e, the eigenvalues of $G$, and not on the particular
dynamics of the system.

The aim of this paper is to apply the synchronization conditions
C1--C3 to a system of two delay-coupled lasers and to laser networks and
discuss the consequences for chaos synchronization. In particular, we
show how condition C1 gives complicated conditions for the coupling
phases of the lasers.

\section{Two Lasers}
\label{sec:twolasers}

\begin{figure}
  \centering
  \includegraphics{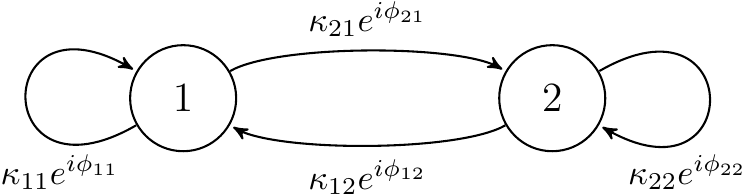}
  \caption{Schematic coupling scheme of two delay-coupled lasers with
    delayed self-feedback. Each connection has a coupling strengths
    $\kappa_{kj}$ and a coupling phase $\phi_{kj}$. We consider the
    case when all delay times are equal.}
  \label{fig:setup}
\end{figure}
In this section we consider two semiconductor lasers that are
delay-coupled to each other with a coupling delay and additionally
receive self-feedback with the same delay time $\tau$. The basic
coupling scheme is depicted in Fig.~\ref{fig:setup}.  The coupled
system is described by dimensionless rate equations of Lang-Kobayashi
type
\begin{eqnarray}
  \dot E_1  &= \shalf (1 +i \alpha)\left[G(n_1,\,E_1) - 1\right] E_1 \nn\\
  &\qquad+ \kappa_{11} e^{i\phi_{11}}E_1(t-\tau) + \kappa_{12}e^{i\phi_{12}} E_2(t-\tau) \;, \nn\\
  \dot E_2  &= \shalf (1 +i \alpha) \left[G(n_2,\,E_2) - 1\right] E_2  \nn\\
  &\qquad+ \kappa_{22}e^{i\phi_{22}} E_2(t-\tau) + \kappa_{21} e^{i\phi_{21}}E_1(t-\tau) \;, \nn\\
  T\dot n_1 &= p - n_1 - G(n_1,\,E_1) |E_1|^2 \;, \nn\\
  T\dot n_2 &= p - n_2 - G(n_2,\,E_2) |E_2|^2 \;,
  \label{eq:laserstart}
\end{eqnarray}
where $E_k$ and $n_k$ are the normalized complex electric field amplitude and the
rescaled inversion of the $k$-th laser, respectively, and $\alpha$ is the linewidth enhancement factor, $p$ is the normalized pump current in excess of the threshold, and
$\kappa_{ij}, \phi_{ij}$
are the coupling amplitudes and phases, respectively, as shown in Fig..~\ref{fig:setup}.   The gain is modeled by
\begin{equation}
  G(n,\,E) = \frac{n+1}{1+\mu|E|^2}
\end{equation}
that takes into account gain saturation effects.  Throughout this
paper we choose the following model parameters for our numerical
simulations, unless stated otherwise: Ratio between carrier and photon
lifetime $T=1000$, $p=0.1$, $\alpha=4$, gain saturation $\mu=0.26$.

The optical coupling phases $\phi_{ij}$ are determined by the optical
path lengths of the feedback and coupling sections on a subwavelength
scale
\begin{equation}
  \phi_{ij} = \Omega_0 \tau_{ij} \;,
\end{equation}
where $\Omega_0$ is the optical frequency of the laser.  Since
$\Omega_0$ is large%
\footnote{Typically, the optical frequency is of the order
  $10^{14}s^{-1}$. We here use dimensionless units, where time is
  measured in units of the photon lifetime (e.g., $T_P=10^{-11}s$). In
  these units typical optical frequencies correspond to a value of
  $\Omega_0\approx 10^{3}$.}, one can for large delay consider the
phases as parameters independent of the coupling delays.  We thus
choose all delays equal, but consider the phases as free parameters.

One important peculiarity of coupled lasers is that the lasers may
synchronize with a relative phase shift $u$
\begin{equation*}
  E_1(t) = e^{iu} E_2(t) \;.
\end{equation*}
To explicitly treat this relative phase shift, we
transform to the new variable $\tilde E_2$ as
\begin{equation}
  \tilde E_2(t) = e^{iu} E_2(t) \;.
\end{equation}
After substituting this into Eqs.~\eqref{eq:laserstart} and omitting the
tildas for simplicity we arrive at the following rate equations for
the fields
\begin{eqnarray}
  \dot E_1  &= \dots \quad+\quad \kappa_{11} e^{i\phi_{11}}E_1(t-\tau) + \kappa_{12}e^{i(\phi_{12}-u)} E_2(t-\tau) \;,\nn\\
  \dot E_2  &= \dots \quad+\quad \kappa_{22}e^{i\phi_{22}} E_2(t-\tau) + \kappa_{21} e^{i(\phi_{21}+u)}E_1(t-\tau) \;.
  \label{eq:main}
\end{eqnarray}
The artificially introduced parameter $u$ helps in the analysis of
synchronization, because we can discuss existence and stability of the
synchronized solution $E_1(t) = E_2(t)$ in dependence on $u$, thereby
treating synchronization of the lasers with a phase shift.

Bringing Eq.~(\ref{eq:main}) into the form of Eq.~(\ref{eq:network})
essentially yields the coupling matrix
\begin{equation}
  G =
  \left[
    \begin{array}{ll}
      \kappa_{11}e^{i\phi_{11}} & \kappa_{12} e^{i(\phi_{12}-u)}\\
      \kappa_{21}e^{i(\phi_{21}+u)} & \kappa_{22} e^{i\phi_{22}}
    \end{array}
  \right]\;.
  \label{eq:matrix}
\end{equation}
The row sum condition C1 is then given by
\begin{equation}
  \kappa_{11} e^{i\phi_{11}} + \kappa_{12}e^{i(\phi_{12}-u)} = \kappa_{22}e^{i\phi_{22}}  + \kappa_{21} e^{i(\phi_{21}+u)} \;.
  \label{eq:phase-condition}
\end{equation}
Equation~\eqref{eq:phase-condition} can be interpreted as follows: If
for a given set of coupling strengths and coupling phases there exists
a phase shift $u$, such that Eq.~\eqref{eq:phase-condition} is
fulfilled, then there exists an invariant synchronization
manifold. The lasers can thus tune their relative phase shifts
appropriately. The stability of the synchronized solution then
determines whether synchronization will be observable or not.

As a starting point for the stability analysis we consider the
\ac{MSF} for a laser network with row sum $\sigma$. Since we focus on
the large delay case ($\tau=1000\gg T_{RO}$) the rotational symmetry
discussed above holds and the \ac{MSF} $\lambda(re^{i\psi})$ depends
solely on $r$.
\begin{figure}
  \centering
  \includegraphics{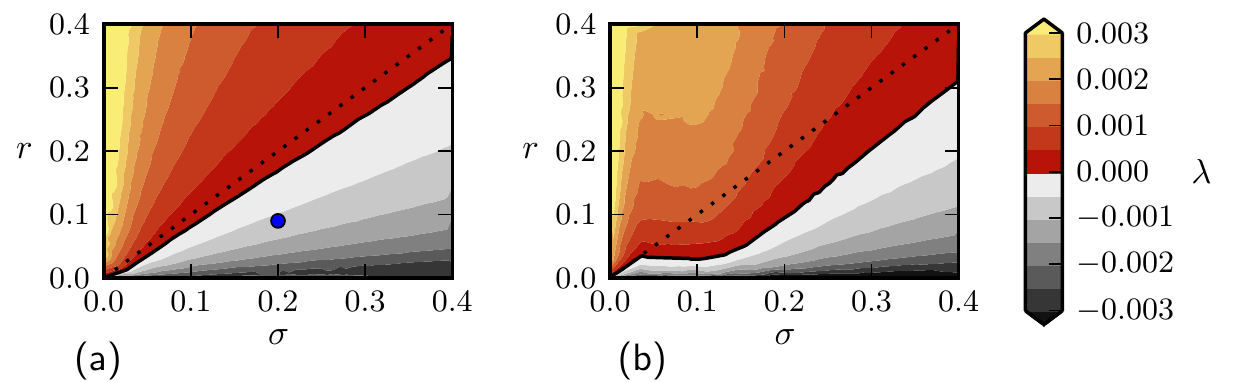}
  \caption{\ac{MSF} $\lambda(r)$ in the $(\sigma,\,r)$-plane
    for $p=0.1$ (panel (a)) and $p=1.0$ (panel (b)).
    The solid line depicts the critical radius $r_0$ with
    $\lambda(r_0)=0$. The dotted line is the diagonal line $r=\sigma$.
    The blue dot corresponds to a parameter set that we will use in the
    following numerical investigations.
    \label{fig:msf}}
\end{figure}
Figure~\ref{fig:msf} depicts the \ac{MSF} in the $(\sigma,\,r)$-plane
for two different values of the pump current $p=0.1$ (panel (a)) and
$p=1$ (panel (b)). When the critical radius $r_0$ (solid line) lies
below the diagonal line $r=\sigma$ (dotted line), the synchronized
solution is chaotic for a network with this row sum, since
$\lambda(\sigma)>\lambda(r_0) = 0$. When $r_0=\sigma$, as occurs for
instance in panel (b) for small values of $\sigma$, the synchronized
dynamics is periodic. In this case the solution $\lambda=0$
corresponds to the Goldstone mode of the periodic orbit.

\begin{figure}
  \centering
  \includegraphics{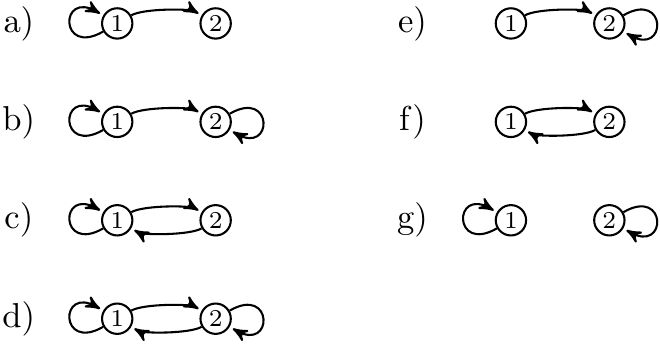}
  \caption{Possible motifs for two delay-coupled lasers. Motifs a)-d)
    can (for certain parameters) exhibit zero-lag chaos
    synchronization and are discussed in more detail below. The motifs
    e)-g) cannot exhibit zero-lag chaos
    synchronization. \label{fig:motifs}}
\end{figure}

In the following we will consider different coupling topologies of the
two lasers. Figure~\ref{fig:motifs} depicts all possible network motifs (up to
exchange of $1\leftrightarrow 2$) with more than one connection. The
motifs on the left (a-d) can exhibit chaos synchronization in the limit of
large delays, while those on the right (e-g) cannot \cite{FLU10b} (trivially
for motif (g)).

We will now discuss the implications of the synchronization conditions
C1--C3 for these motifs and compare the
predictions with numerical simulations. We study two measures for
synchronization. The first measure is the correlation coefficient of
the laser intensities $I_1, I_2$
\begin{equation}
  \rho=\frac{\Bigl\langle (I_1-\langle I_1\rangle)\,(I_2 -\langle I_2\rangle)\Bigr\rangle}{\sqrt{(\Delta I_1)^2(\Delta I_2)^2}} \;,
\end{equation}
where $\langle \cdot \rangle$ denotes the time average and $(\Delta
I_k)^2$ denotes the variance of the respective intensity.  Although,
the correlation coefficient can in principle distinguish between
identical synchronization ($\rho=1$) and imperfect (generalized)
correlation ($\rho<1$), it is not the most sensitive measure for this
purpose, because imperfect synchronization may still yield very large
correlation $\rho\approx 1$.

To overcome this disadvantage of the correlation coefficient, we
calculate the synchronization probability $P_S$.  We define $P_S$ as
the probability that at any time $t$ the relative error between
$I_1(t)$ and $I_2(t)$ is smaller than a threshold $\varepsilon$
\begin{equation}
  P_S = \mathrm{Prob}\left(\frac{|I_1(t)-I_2(t)|}{\langle I_1\rangle +\langle I_2\rangle} < \varepsilon \right)\;.
\end{equation}
We choose $\varepsilon=0.01$ in the following.

\subsection{Motif (a)}
This case is the classical master slave configuration for chaos
communication with lasers.  It is also referred to as open-loop master
slave configuration \cite{VIC02}, since the receiver (2) has no
self-feedback.  The coupling matrix Eq.~\eqref{eq:matrix} is in this
case given by
\begin{equation*}
  G =
  \left[
    \begin{array}{ll}
      \kappa_{11}e^{i\phi_{11}} & 0 \\
      \kappa_{21}e^{i(\phi_{21}+u)} & 0
    \end{array}
  \right]\;.
\end{equation*}
The row sum condition C1 then becomes
\begin{equation*}
  \kappa_{11}e^{i\phi_{11}} = \kappa_{21} e^{i(\phi_{21} + u)} \;.
\end{equation*}
This condition is satisfied if and only if
$\kappa_{11}=\kappa_{21}$. The phase shift $u$ can then compensate any
choice of coupling phases $\phi_{11}$ and $\phi_{21}$. The second
eigenvalue of the coupling matrix is zero, such that the motif is
optimal for chaos synchronization and condition C2 is always
fulfilled.  Whether the systems will synchronize or not then depends
on whether $\lambda(0)$ is smaller or larger than zero, i.e., whether
the \ac{MSF} in Fig.~\ref{fig:msf} is negative (weak chaos) or
positive (strong chaos) for $r=0$.

Note that since the coupling is unidirectional, the synchronization
properties do not depend on the coupling delay. In particular, if the
coupling delay is smaller than the self-feedback delay, anticipated
synchronization \cite{VOS00} can be observed \cite{MAS01}.

\subsection{Motif (b)}
This setup consists of two unidirectionally coupled lasers with
self-feedback and has been studied in different contexts.  The
importance of the coupling phases in this coupling scheme has been
recognized in \cite{PEI02}.  In this reference it was observed in an
experiment that depending on the (relative) feedback phases the
synchronization behavior ranges from perfect synchronization to an
almost uncorrelated state. So far these experiments have not been
sufficiently explained.  It turns out that the experimental results
can be well understood in the light of the synchronization conditions
C1--C3:

The coupling matrix corresponding to motif (b) is given by
\begin{equation}
  G =
  \left[
    \begin{array}{ll}
      \kappa_{11}e^{i\phi_{11}} & 0 \\
      \kappa_{21}e^{i(\phi_{21}+u)} & \kappa_{22} e^{i\phi_{22}}
    \end{array}
  \right]
\end{equation}
and the row sum condition C1 reads
\begin{equation*}
  \kappa_{11} e^{i\phi_{11}} = \kappa_{21} e^{i(\phi_{21} + u)} + \kappa_{22}e^{i\phi_{22}} \;.
\end{equation*}
Eliminating $u$, yields the following condition on the coupling strengths and phases
\begin{equation}
  \kappa_{21}=\left|\kappa_{11}-\kappa_{22} e^{i\Phi_{\rm rel}}\right| = \sqrt{\kappa_{11}^2 + \kappa_{22}^2 - 2\kappa_{11} \kappa_{22}\cos(\Phi_{\rm rel})} \label{eq:motifbC1}
\end{equation}
with $\Phi_{\rm rel} = \phi_{22}-\phi_{11}$. The phase shift $u$
between the laser fields is then given by
\begin{equation}
  u = \phi_{11}-\phi_{21} + \Arg\left(\kappa_{11}-\kappa_{22} e^{i\Phi_{\rm rel}}\right) \;,
\end{equation}
where $\Arg$ denotes the complex argument.

Condition C2 on the other hand reads for this case
\begin{equation*}
  \kappa_{22} < \kappa_{11} = |\sigma| \;,
\end{equation*}
i.e., chaos synchronization is possible, if the self-coupling of
laser~2 is weaker than that of laser~1.  For a given set of
parameters, synchronization is stable if condition C3 is fulfilled,
i.e., if
\begin{equation}
  \kappa_{22} < r_0(\kappa_{11}) \;.
\end{equation}

To illustrate these conditions, we consider the following parameter
set
\begin{equation}
  \kappa_{11}=0.2,\quad\kappa_{22}=0.09,\qquad\phi_{11}=0,\qquad\phi_{21}=0
\end{equation}
and vary $\Phi_{\rm rel} = \phi_{22}$ and $\kappa_{21}$. These
coupling strengths $(\sigma,\,r)=(\kappa_{11},\,\kappa_{22})$ are
marked by a blue dot in Fig.~\ref{fig:msf}a and are chosen such that
synchronization is stable (condition C3 is fulfilled).
\begin{figure}
  \centering
  \includegraphics{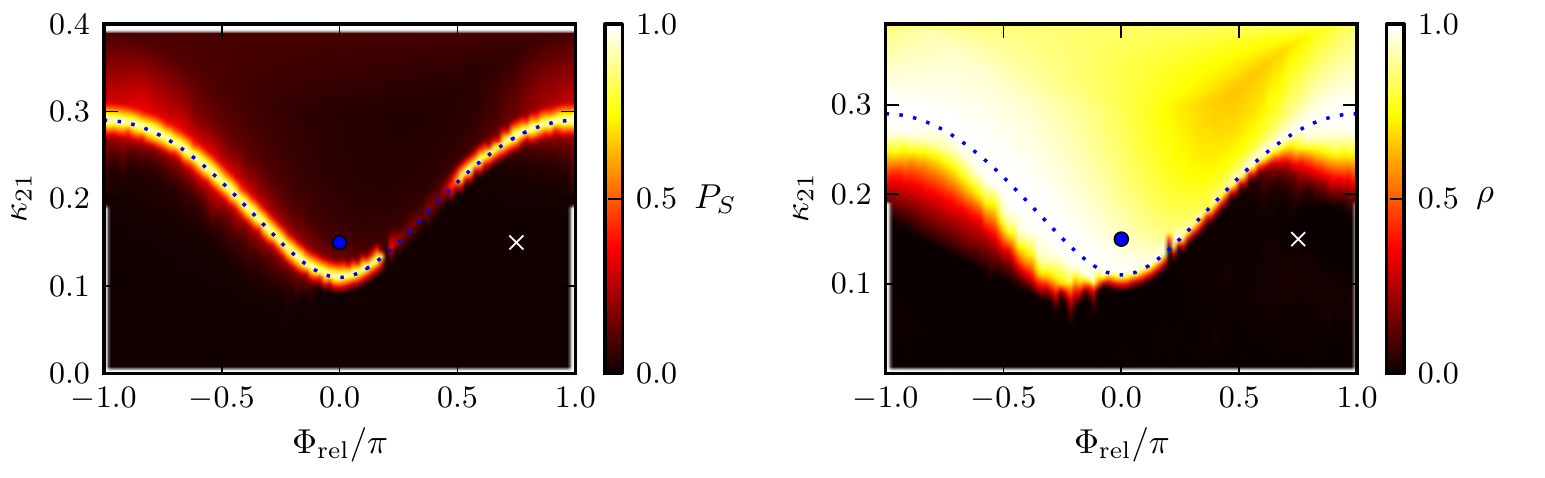}
  \caption{\label{fig:motifb} (Motif (b)) Synchronization probability
    $P_S$ (left) and correlation coefficient $\rho$ (right) as a function of $\phi_{22}         =\Phi_{\rm rel}$ and $\kappa_{21}$. The dotted line corresponds to the
    synchronization condition C1 (Eq.~(\ref{eq:motifbC1})). Other
    parameters: $\kappa_{11}=0.2$, $\kappa_{22}=0.09$,
    $\phi_{11}=\phi_{21}=0$.}
\end{figure}
Figure~\ref{fig:motifb} depicts the correlation coefficient $\rho$ and
the synchronization probability $P_S$ in the $(\Phi_{\rm rel},\,
\kappa_{21})$-plane. The dotted line corresponds to
Eq.~(\ref{eq:motifbC1}), where condition C1 is satisfied. This
condition clearly coincides with high synchronization probabilities.

\begin{figure}
  \centering
  \includegraphics{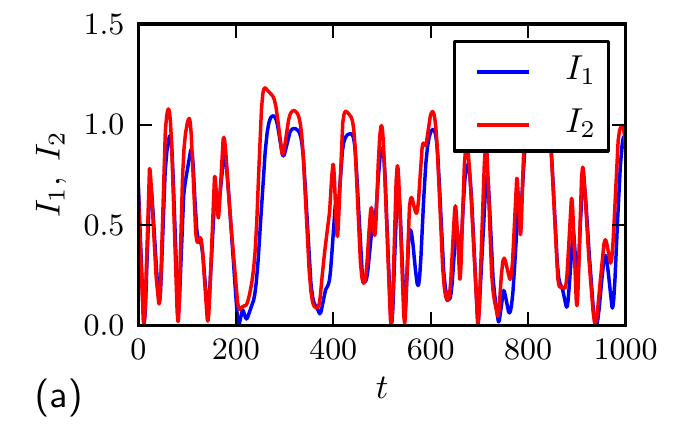}
  \includegraphics{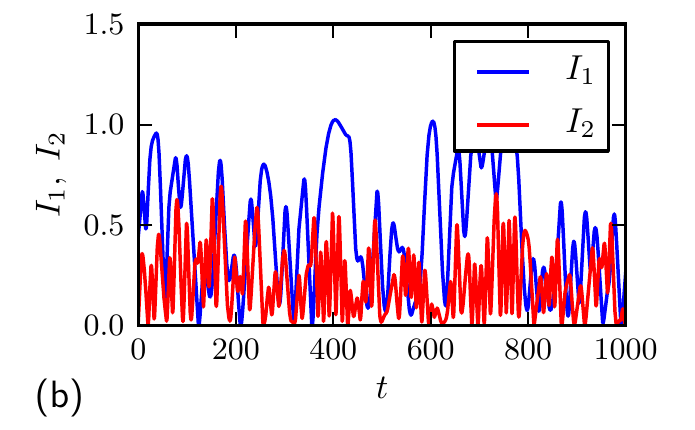}
  \caption{\label{fig:generalized} Panel (a): Generalized
    synchronization for $\kappa_{21}=0.15$ $\phi_{22}=\Phi_{\rm
      rel}=0$ (blue dot in Fig.~\ref{fig:motifb}).  Panel (b): No
    synchronization for $\kappa_{21}=0.15$ $\phi_{22}=\Phi_{\rm
      rel}=\frac{3}{4}\pi$ (white cross in Fig.~\ref{fig:motifb}).
    Other parameters as in Fig.~\ref{fig:motifb}.}
\end{figure}

As discussed above, high correlations are also possible without
perfect synchronization. In the regions of high correlation and low
synchronization probabilities, we observe generalized synchronization.
Figure~\ref{fig:generalized}(a) shows an example intensity time series
observed in these regions. The parameters correspond to the blue dot
in Fig~\ref{fig:motifb}.
The blue dotted line (condition C1) in Fig~\ref{fig:motifb} roughly
marks the boundary between regions of high and low correlations. Below
the line, we observe no synchronization at all. An examplary time
series in this regime (corresponding to the white cross in
Fig.~\ref{fig:motifb}) is shown in Fig.~\ref{fig:generalized}(b).  The
correlation in this case is very low. Interestingly, the dynamics of
the second laser has, in this case, a strong high-frequency
component. The occurrence of this high-frequency component can be
understood as a result of the non-locking behavior. Since laser~2 does
not lock to the signal of laser~1, the overall input signal of
laser~2 is given by the interference of two (almost) independent
chaotic signals, namely the signal from laser~1 and the self-feedback
signal from laser~2. In this interference signal the high-frequency
component is present due to a fast alternation of constructive and
destructive interference. Indeed, by switching off the coupling
($\kappa_{21}=0$) and calculating the intensity of an interference
signal $I_{\rm intf} = |E_1(t) + E_2(t)|^2$ one can already observe
the high-frequency component. For non-zero coupling this interference
signal drives laser~2, leading to even higher-order effects.

In Fig.~\ref{fig:motifb}(left) there is a region on the dotted line
$\Phi_{\rm rel} \in [0.2\pi, 0.5\pi]$, where the synchronization
probability is low. In this region we observe multistability between
identical synchronization solution, a state of generalized
synchronization similar to Fig.~\ref{fig:generalized}(a), and an
uncorrelated state similar to Fig.~\ref{fig:generalized}(b). Which
state is chosen depends in our deterministic simulations sensitively
on the initial conditions. Including noise in the simulation results
in spontaneous switching between the three states, albeit it is hard
to distinguish between identical synchronization and generalized
synchronization in the presence of noise.

Thus although our synchronization condition C1--C3 give the correct
existence and stability of the identical synchronized solution, we
certainly cannot exclude the existence of other attractors.

The experimental investigations in Ref.~\cite{PEI02} were performed
under similar parameter conditions, i.e., $\kappa_{22}<\kappa_{11}$
(obeying condition C2).  As $\Phi_{\rm rel}$ was varied, the
correlation varied from almost perfect to almost no correlation.
Varying $\Phi_{\rm rel}$ for a fixed value of $\kappa_{21}$ in
Fig.~\ref{fig:motifb} reproduces this behavior, provided the value of
$\kappa_{21}$ intersects the dotted synchronization curve.  Thus the
experimental results can be understood as interference effects, which
may or may not lead to the existence of a synchronization manifold,
corresponding to high and low correlation, respectively.

\subsection{Motif (c)}
For this motif the coupling matrix is given by
\begin{equation}
  G =
  \left[
    \begin{array}{ll}
      \kappa_{11}e^{i\phi_{11}} & \kappa_{12} e^{i(\phi_{12}-u)}\\
      \kappa_{21}e^{i(\phi_{21}+u)} & 0\\
    \end{array}
  \right]
  \label{eq:motifc:G}
\end{equation}
and the row sum condition C1 reads
\begin{equation}
  \kappa_{11}e^{i\phi_{11}} + \kappa_{12} e^{i(\phi_{12}-u)} = \kappa_{21}e^{i(\phi_{21}+u)} \;. \label{eq:motifcC1}
\end{equation}
In an experiment with a simple bi-directional coupling, i.e.,
face-to-face coupling of the lasers, the coupling from laser~1 to
laser~2 has the same coupling strengths as the reverse direction:
\begin{equation}
  \kappa_{12} \approx  \kappa_{21} \;. \label{eq:motifc:natural}
\end{equation}
We will first consider the more general case, which can be realized in
an experiment using optical isolators and separate beam paths for the
two coupling directions. We will then consider the natural condition
Eq.~\eqref{eq:motifc:natural} as a special case.

Our aim now is to derive a condition that is equivalent to the
existence of a phase shift $u$ satisfying Eq.~(\ref{eq:motifcC1}),
i.e., we want to eliminate $u$ from the equation.  To simplify the
discussion, we introduce two parameters:
\begin{equation}
  v=u + \frac{\phi_{21}-\phi_{12}}{2},\quad\mbox{and}\quad \theta=\frac{\phi_{21}+\phi_{12}}{2} \;. \label{eq:motifc:parameters}
\end{equation}
Note that $v$ is a ``free'' parameter, since the phase shift $u$
can be selected by the system.  With these parameters
Eq.~(\ref{eq:motifcC1}) can be written as
\begin{equation}
  \kappa_{11}e^{i(\phi_{11}-\theta)} = -\kappa_{12} e^{-iv} + \kappa_{21} e^{iv} \;.\label{eq:motifc:e1}
\end{equation}
For varying $v$ the terms on the right hand side describe an ellipse
in the complex plane with semi-minor axis
$a=|\kappa_{12}-\kappa_{21}|$ oriented along the real axis and
semi-major axis $b=|\kappa_{12}+\kappa_{21}|$ oriented along the
imaginary axis. In order for Eq.~(\ref{eq:motifc:e1}) to have a
solution, the left hand side has to lie on this ellipse. Thus
\begin{eqnarray}
  x:=\mathrm{Re}\Bigl(\kappa_{11}e^{i(\phi_{11}-\theta)}\Bigr)=\kappa_{11} \cos(\phi_{11}-\theta) \;,\\
  y:=\mathrm{Im}\Bigl(\kappa_{11}e^{i(\phi_{11}-\theta)}\Bigr)=\kappa_{11} \sin(\phi_{11}-\theta)
\end{eqnarray}
have to obey
\begin{equation}
  \frac{x^2}{a^2} + \frac{y^2}{b^2} = 1 \;. \label{eq:motifc:finalC1}
\end{equation}
\begin{SCfigure}[][tb]
  \includegraphics{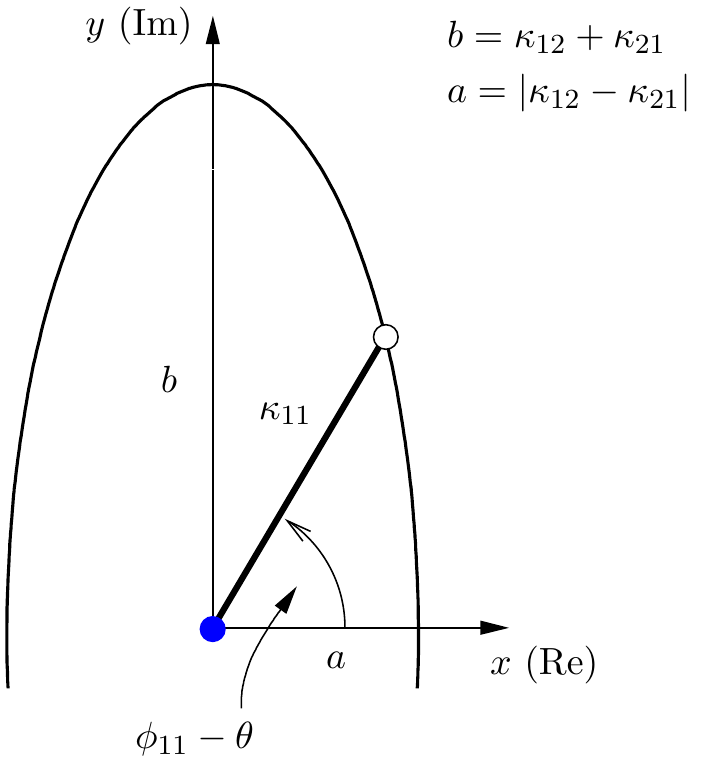}
  \centering
  \caption[Geometric visualization of the synchronization condition.]
  {\label{fig:mechanical_c} (Motif (c)) Geometric visualization of the
    synchronization condition (Eq.~(\ref{eq:motifc:e1})).  If the
    complex number $\kappa_{11}e^{i(\phi_{11}-\theta)}$ lies on the
    ellipse, there exists an invariant synchronization manifold.}
\end{SCfigure}
\begin{figure}[tb]
  \centering
  \includegraphics{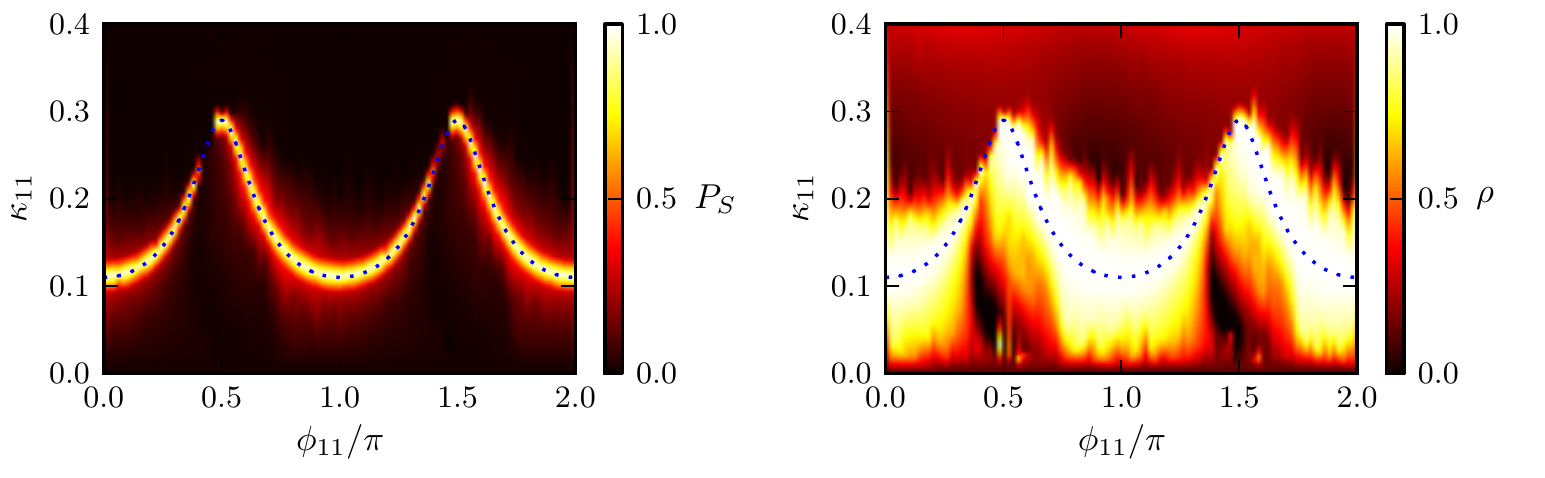}
  \caption{\label{fig:motifc} (Motif (c)) Synchronization probability
    $P_S$ (left) and correlation coefficient $\rho$ (right) as a function of $\phi_{11}$     and
    $\kappa_{11}$. The dotted line corresponds to the synchronization
    condition C1 (Eq.~(\ref{eq:motifc:k11})). Other parameters:
    $\kappa_{21}=0.2$, $\kappa_{12}=0.09$, $\phi_{12}=\phi_{21}=0$.}
\end{figure}
This equation is the desired synchronization condition corresponding
to C1. It involves the coupling strengths $\kappa_{11}$,
$\kappa_{12}$, and $\kappa_{21}$ and the coupling phases $\phi_{11}$,
$\phi_{12}$, and $\phi_{21}$. We can write it as an explicit equation
for $\kappa_{11}$
\begin{equation}
  \kappa_{11}=\Bigl[ \frac{\cos^2(\phi_{11}-\theta)}{(\kappa_{12}-\kappa_{21})^2} + \frac{\sin^2(\phi_{11}-\theta)}{(\kappa_{12}+\kappa_{21})^2}  \Bigr]^{-1/2} \;.
  \label{eq:motifc:k11}
\end{equation}
Note that for the degenerate case $\kappa_{12}=\kappa_{21}$, i.e.,
$a=0$, the relevant condition becomes
\begin{equation}
  x=0,\qquad\mbox{and}\qquad y^2\le b^2\;.
\end{equation}

A necessary condition on the coupling strengths for the existence of
a solution is
\begin{equation}
  |\kappa_{12}-\kappa_{21}|  \le \kappa_{11} \le \kappa_{12}+\kappa_{21} \;.
\end{equation}
To obtain the phase shift $u$, we split Eq.~(\ref{eq:motifc:e1}) into real
and imaginary part and solve for $v$. This yields
\begin{equation}
  v = \Arg\left(\frac{x}{\kappa_{21}-\kappa_{12}} + i  \,\frac{y}{\kappa_{21}+\kappa_{12}} \right) \;.
\end{equation}
The phase shift $u$ is obtained from the definition of
$v=u+(\phi_{21}-\phi_{12})/2$.

We now discuss condition C2 and C3, which concern the eigenvalue of
$G$ other than the row sum. When $G$ has a constant (complex) row sum
$\sigma$, this row sum is an eigenvalue of $G$. Since the determinant
of a matrix is the product of its eigenvalues, we have the following
equation for the second eigenvalue $\gamma_1$ of $G$
\begin{equation}
  |\det G| = |\kappa_{12} \kappa_{21}|=|\sigma| |\gamma_1|\;.
\end{equation}
On the other hand, there is only one entry
$\kappa_{21}e^{i(\phi_{21}+u)}$ in the second row of $G$ (see
Eq.~(\ref{eq:motifc:G})), such that in the case of constant row
sum we have
\begin{equation}
  |\sigma| = \kappa_{21},\qquad\mbox{and}\qquad |\gamma_1| = \kappa_{12}\;.
\end{equation}
Thus chaos synchronization is possible if (C2)
\begin{equation}
  \kappa_{12}<\kappa_{21} = |\sigma|\;. \label{eq:motifc:C2}
\end{equation}
For given parameters (given $r_0$) synchronization is stable if and
only if (C3)
\begin{equation}
  \kappa_{12} < r_0(\kappa_{21})\;.
\end{equation}

Again, we illustrate these conditions by numerical simulations. We choose
\begin{equation}
  \kappa_{12}=0.09,\qquad\kappa_{21}=0.2,\qquad \phi_{12}=0,\qquad \phi_{21}=0\;.
\end{equation}
The coupling strength again corresponds to the blue dot in
Fig.~\ref{fig:msf}a, such that synchronization is stable if a
synchronization manifold exists.  Similarly to the case of motif (b),
there are small regions of low synchronization probability on the
dotted curve in Fig.~\ref{fig:motifc}(left). These again correspond to
regions of multistability and the simulations depend sensitively on
initial conditions.

Coming back to the most natural face-to-face coupling
(Eq.~\eqref{eq:motifc:natural}) with
\begin{equation}
  \kappa_{12}=\kappa_{21}
\end{equation}
it becomes clear from Eq.~(\ref{eq:motifc:C2}) that chaos
synchronization is not possible in this simple setup. To synchronize
the two lasers in this motif one needs an asymmetric coupling.

\subsection{Motif (d)}
We now consider motif (d), where all possible couplings are
present.  The corresponding coupling matrix is given by
\begin{equation}
  G =
  \left[
    \begin{array}{ll}
      \kappa_{11}e^{i\phi_{11}} & \kappa_{12} e^{i(\phi_{12}-u)}\\
      \kappa_{21}e^{i(\phi_{21}+u)} & \kappa_{22}e^{i\phi_{22}}
    \end{array}
  \right] \;.
  \label{eq:motifd:G}
\end{equation}
The row sum condition C1 is given by
\begin{equation}
  \kappa_{11}e^{i\phi_{11}}  + \kappa_{12} e^{i(\phi_{12}-u)} =  \kappa_{21} e^{i(\phi_{21}+u)} + \kappa_{22} e^{i\phi_{22}} \;. \label{eq:motifdC1}
\end{equation}
Below we will discuss this condition in more detail. First, however,
we turn to conditions C2 and C3 involving the eigenvalues $\sigma$
(row sum) and $\gamma$ of the matrix $G$.

Assuming that the row sum condition Eq.~(\ref{eq:motifdC1}) is
satisfied, we solve Eq.~(\ref{eq:motifdC1}) for
$\kappa_{11}e^{i\phi_{11}}$ and replace this term in the coupling
matrix~(\ref{eq:motifd:G}). The resulting matrix then has the
eigenvalues
\begin{eqnarray}
  \sigma &= \kappa_{22} e^{i\phi_{22}} + \kappa_{21}e^{i(\phi_{21}+u)} \qquad \mbox{and}\\
  \gamma &= \kappa_{22} e^{i\phi_{22}} - \kappa_{12}e^{i(\phi_{12}-u)} \;.
\end{eqnarray}
The condition C2 ($|\gamma|<|\sigma|$) gives (after some straightforward
calculations)
\begin{eqnarray}
  \kappa_{12}^2 &-2\kappa_{12}\kappa_{22}\cos(\phi_{12} - \phi_{22}-u) \nonumber\\
  & < \kappa_{21}^2 + 2\kappa_{21}\kappa_{22} \cos(\phi_{21}-\phi_{22}+u) \;. \label{eq:motifd:C2}
\end{eqnarray}
The sufficient condition C3 similarly yields
\begin{equation}
  \sqrt{\kappa_{12}^2 - 2\kappa_{12}\kappa_{22} \cos(\phi_{12}-\phi_{22}-u) + \kappa_{22}^2} < r_0(\sigma) \;.
  \label{eq:motifd:C3}
\end{equation}
In contrast to the previous cases, it is not possible to eliminate the phase shift $u$
between the lasers directly. We will, however, obtain a formula for $u$ below.

\begin{SCfigure}[][tb]
  \includegraphics{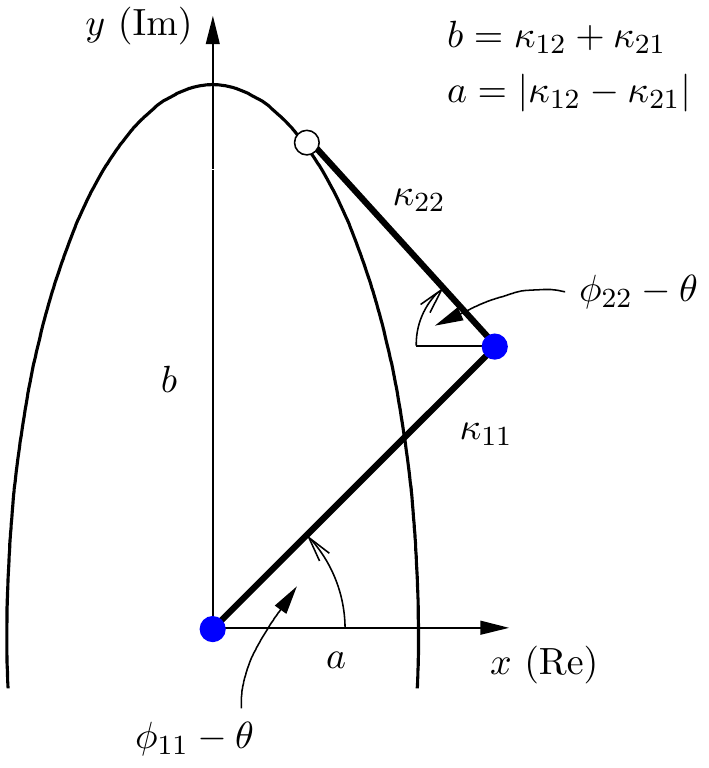}
  \centering
  \caption[Geometric visualization of the synchronization condition.]
  {\label{fig:mechanical} (Motif (d)) Geometric visualization of the
    synchronization condition (Eq.~(\ref{eq:motifd:e1})).  If the sum
    of the two complex numbers $\kappa_{11}e^{i(\phi_{11}-\theta)}$
    and $\kappa_{22}e^{i(\phi_{22}-\theta)}$ lies on the ellipse,
    there exists an invariant synchronization manifold.}
\end{SCfigure}

We now further discuss the row sum condition Eq.~(\ref{eq:motifdC1})
and aim to eliminate the relative phase shift $u$. The calculation is
similar to that done for motif (c), only more involved. Using the same
parameters as for the case of motif (c)
(Eq.~(\ref{eq:motifc:parameters}))
\begin{equation}
  v=u + \frac{\phi_{21}-\phi_{12}}{2},\quad\mbox{and}\quad \theta=\frac{\phi_{21}+\phi_{12}}{2} \;,
\end{equation}
we obtain
\begin{equation}
  \label{eq:motifd:e1}
  \kappa_{11}e^{i(\phi_{11}-\theta)} -\kappa_{22} e^{i(\phi_{22}-\theta)}
  = -\kappa_{12} e^{-i v} + \kappa_{21} e^{i v} \;.
\end{equation}
Again $v$ is a free parameter, since it is proportional to the
relative phase shift $u$, which can be selected by the system.  This
equation corresponds to the geometric problem visualized in
Fig.~\ref{fig:mechanical}.  For varying $v$ the terms on the right
hand side describe an ellipse in the complex plane with semi-minor
axis $a=|\kappa_{12}-\kappa_{21}|$ oriented along the real axis and
semi-major axis $b=\kappa_{12}+\kappa_{21}$ oriented along the
imaginary axis.  For the equation to have a solution the real and
imaginary part of the left hand side have to lie on this ellipse.
Thus
\begin{eqnarray}
  x &:=
  \kappa_{11}\cos(\phi_{11}-\theta) - \kappa_{22}\cos(\phi_{22}-\theta) \nn \\
  y &:=
  \kappa_{11}\sin(\phi_{11}-\theta) - \kappa_{22}\sin(\phi_{22}-\theta) \label{eq:motifd:xandy}
\end{eqnarray}
have to obey
\begin{equation}
  \frac{x^2}{a^2} + \frac{y^2}{b^2} = 1 \;. \label{eq:motifd:finalC1}
\end{equation}
Equation (\ref{eq:motifd:finalC1}) is the final condition, which has
to be fulfilled in order for the synchronization manifold to be invariant.

\begin{figure}[tb]
  \centering
  \includegraphics{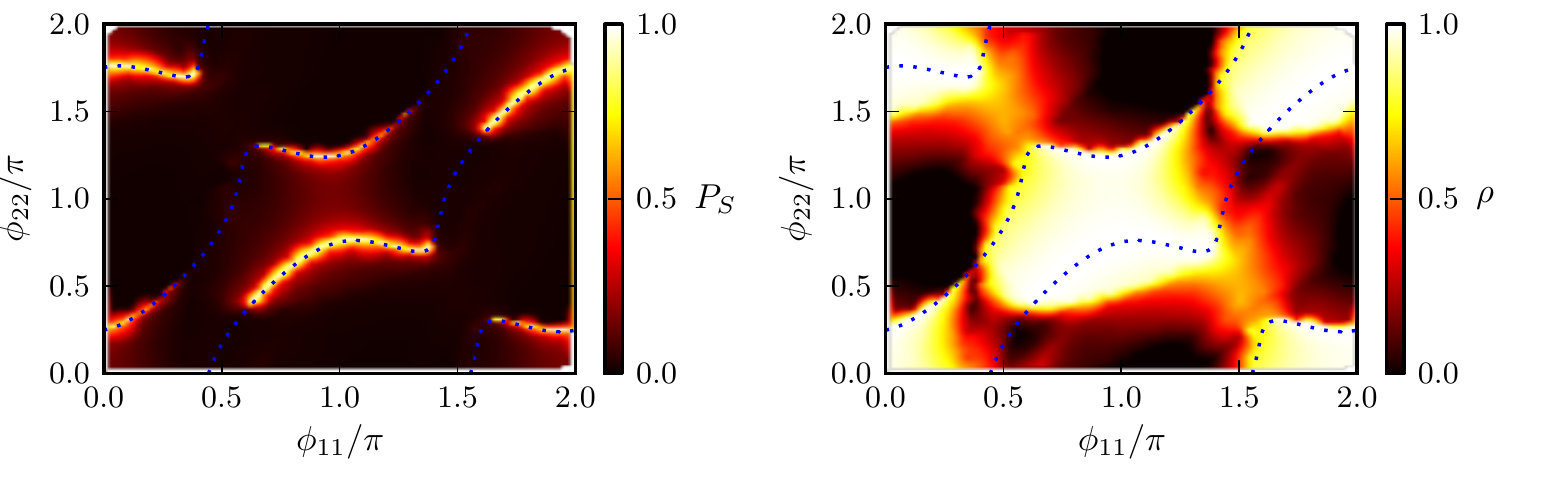}
  \caption{\label{fig:motifd} (Motif (d)) Synchronization probability
    $P_S$ (left) and correlation coefficient $\rho$ (right) as a function of $\phi_{11}$     and
    $\phi_{22}$. The dotted line corresponds to the synchronization
    condition C1 (Eq.~(\ref{eq:motifd:finalC1})).  Other parameters:
    $\kappa_{11}=0.25$, $\kappa_{12}=0.1$, $\kappa_{21}=0.25$,
    $\kappa_{22}=0.15$, $\phi_{12}=\phi_{21}=0$.}
\end{figure}

Movie~1 (\texttt{animation.mpg}) illustrates how the geometrical
problem of Fig.~\ref{fig:mechanical} results in the dotted
synchronization curves in Fig.~\ref{fig:motifd}.

Figure~\ref{fig:motifd} depicts the correlation coefficient and the
synchronization probability in the ($\phi_{11},\, \phi_{22})$-plane
for a fixed set of coupling strengths and cross-coupling phases.  On
the dotted lines the synchronization condition C1
(Eq.~(\ref{eq:motifd:finalC1})) is fulfilled. The lasers exhibit
strong synchronization on this curve.  For this motif, the row sum
$\sigma$ and the transversal eigenvalue $\gamma$ depend on the
parameters $\phi_{11}$ and $\phi_{22}$. Due to this dependence, the
synchronization condition C3 is not satisfied everywhere on the dotted
line and identical synchronization is only stable on parts of the
curve.  Additionally, we again have the effect of multistability as
discussed for the last two motifs, which can result in low
synchronization probability on the dotted curve.

Similarly to the case of motif (c), the relative phase shift $u$ can
be found by solving Eq.~(\ref{eq:motifd:e1}) for $v$
\begin{equation}
  v = \Arg\left(\frac{x}{\kappa_{21}-\kappa_{12}} + i  \,\frac{y}{\kappa_{21}+\kappa_{12}} \right) \;
  \label{eq:tanphi}
\end{equation}
and using the definition of $v=u+(\phi_{21}-\phi_{12})/2$. This phase
shift can then be used in condition C2 and C3
(Eqs.~(\ref{eq:motifd:C2}) and (\ref{eq:motifd:C3})).

In order for Eq.~(\ref{eq:motifd:finalC1}) to have a solution, the two
vectors with respective lengths $\kappa_{11}$ and $\kappa_{22}$ have
to be able to reach the ellipse, \ie, the sum of the magnitudes
$\kappa_{11}+\kappa_{22}$ has to be larger or equal to the length of
the semi-minor axis $a=|\kappa_{12}-\kappa_{21}|$. Similarly, the
absolute value of the magnitude difference $|\kappa_{11}-\kappa_{22}|$
has to be smaller or equal to the length of the semi-major axis
$b=\kappa_{12} + \kappa_{21}$. We thus obtain two conditions for the
existence of a solution
\begin{eqnarray}
  (\kappa_{11}+\kappa_{22})^2 &\ge& (\kappa_{12}-\kappa_{21})^2 \;, \nn\\
  (\kappa_{11}-\kappa_{22})^2 &\le& (\kappa_{12}+\kappa_{21})^2 \;.
  \label{eq:motifd:kcond}
\end{eqnarray}
If and only if the coupling strengths fulfill
Eqs.~(\ref{eq:motifd:kcond}), there is a combination of phases such
that condition (\ref{eq:motifd:finalC1}) is satisfied.

In many optical setups the forward and backward directions have
approximately equal coupling strengths
$\kappa_{12}\approx\kappa_{21}$.  This holds for instance for a
setup where the lasers are coupled via a common mirror (see
Sec.~\ref{sec:commonmirror}) or where the lasers are coupled
face-to-face.  In this case the ellipse becomes a line along the
$y$-axis stretching from $-b$ to $b$. Thus
Eq.~(\ref{eq:motifd:finalC1}) reduces to $x=0$ with the
supplementary condition $y^2\le b^2$.

\section{Laser networks}
\label{sec:networks}

We now move from the two-laser system to networks of $N$ all-optically
coupled lasers. The synchronization of laser networks is on one hand
important for applications for instance in high power laser arrays,
where the synchronization of optical phases yields an intensity
$I\propto N^2$ for interfering beams in contrast to the case of
randomly distributed phases that gives $I\propto N$
\cite{KOZ00,WIE09a}.

On the other hand, laser networks have been proposed as optical
information processing systems \cite{AMA08,APP11}. Understanding
stability properties of dynamical states in these networks is a
necessary first step for controlling and utilizing these systems.

For a network of $N$ all-optically coupled lasers, the coupling terms
in the Lang-Kobayashi rate equations are given by
\begin{eqnarray}
  \dot E_k  = \dots + \sum_{j=1}^N \kappa_{kj}\, e^{i\phi_{kj}} E_j(t-\tau) \;.
  \label{eq:lasernetwork}
\end{eqnarray}
Here $\kappa_{kj}$ and $\phi_{kj}$ are matrices describing the
coupling strength and coupling phase, respectively, of the connection
$j\rightarrow k$.  Again the lasers may synchronize with relative
phase shifts $u_k$, which we define with respect to laser~1
\begin{equation}
  E_1(t) = e^{iu_k} E_k(t) \;,
\end{equation}
i.e., $u_1=0$.
Performing the corresponding transformation (as before in Eq.~\eqref{eq:main})
\begin{equation}
  \tilde{E}_k(t) = e^{iu_k} E_k(t)
\end{equation}
and omitting the tildas for simplicity we obtain the field equations
\begin{equation}
  \dot E_k  = \dots + \sum_{j=1}^N \kappa_{kj}\,e^{i(\phi_{kj} + u_k-u_j)} E_j(t-\tau) \;.
\end{equation}
The necessary condition C1 for synchronization is then stated as
follows: There exists an invariant synchronization manifold if and only if there is a
combination of relative phases $u_k$ ($k=2,\dots,N$), such that the
complex row sum
\begin{equation}
  \sigma_k = \sum_{j=1}^N \kappa_{kj} \, e^{i(\phi_{kj}+u_k-u_j)}
\end{equation}
is independent of $k$.  As we saw before, this condition is already
very difficult to analyze for two lasers when all four connections are
present. In an experiment it may be possible to actively control two
or three phases accurately using for instance piezo positioning of
mirrors, or passive wave-guides where the optical path-lengths can be
controlled through an injection current \cite{SCH06a}.  However, it is
virtually impossible to actively control more than a few phases in
this way. For larger networks, optoelectronic
\cite{UDA01,CAL09,RAV11}, incoherent optical feedback
\cite{ROG01b,SUK04} or coupling via a common relay \cite{FIS06,ZAM10}
may thus be a more promising coupling method. However, in certain
cases all-optical coupling may be feasible in an experimental
situation. We discuss such a feasible setup below.

To illustrate the complexity of the synchronization condition C1 in a
simple case, we consider a system of bidirectionally coupled lasers.

\subsection{Rings of bi-directionally coupled lasers}

For rings of bi-directionally coupled lasers, the coupling matrix
has the following principal structure
\newcommand{\zn}{0\phantom{G_{12}}}
\begin{equation}
  G =
  \left[
    \begin{array}{ccccc}
      0     & G_{12} &        &          & G_{1N}\\
      G_{21} & 0     & G_{23}  &          & \\
      & G_{32} & \ddots & \ddots   & \\
      &       & \ddots & \ddots   & G_{N-1,N} \\
      G_{N1} &       &        & G_{N,N-1} & 0
    \end{array}
  \right] \;,
\end{equation}
with $G_{kj} = \kappa_{kj} \, e^{i(\phi_{kj} + u_k - u_j)}$.
We then obtain the following row sums
\begin{eqnarray}
  \sigma_1 &= \kappa_{12} \,e^{i(\phi_{12} + u_1 - u_2)} + \kappa_{1N} \,e^{i(\phi_{1N} + u_1 - u_N)} \;,\nn\\
  \sigma_2 &= \kappa_{23} \,e^{i(\phi_{23} + u_2 - u_3)} + \kappa_{21} \,e^{i(\phi_{21} + u_2 - u_1)} \;,\nn\\
  \vdots   & \nn\\
  \sigma_{N-1} &= \kappa_{N-1,N} \,e^{i(\phi_{N-1,N} + u_{N-1} - u_N)} + \kappa_{N-1,N-2} \,e^{i(\phi_{N-1,N-2} + u_{N-1} - u_{N-2})} \;,\nn\\
  \sigma_{N} &= \kappa_{N,1} \,e^{i(\phi_{N,1} + u_{N} - u_1)} + \kappa_{N,N-1} \,e^{i(\phi_{N,N-1} + u_{N} - u_{N-1})} \;. \label{eq:ring-rowsums}
\end{eqnarray}
The necessary synchronization condition C1 then corresponds to
$2(N-1)$ equations (each complex equation yields two real equations)
\begin{equation}
  \sigma_1 = \sigma_2,\quad \sigma_2=\sigma_3,\quad\dots\quad \sigma_{N-1}=\sigma_N\;.
\end{equation}
However, there are only $N-1$ relative phases $u_j$ ($u_1=0$) that the
system can choose freely. Thus, in an experiment we need to control
$N-1$ coupling phases to satisfy the synchronization condition C1.
For a ring of $N=3$ elements, controlling two coupling phases may
still be feasible, but this approach quickly fails for larger $N$.

\subsection{Coupling via a common mirror}
\label{sec:commonmirror}
\begin{SCfigure}[][tb]
  \includegraphics{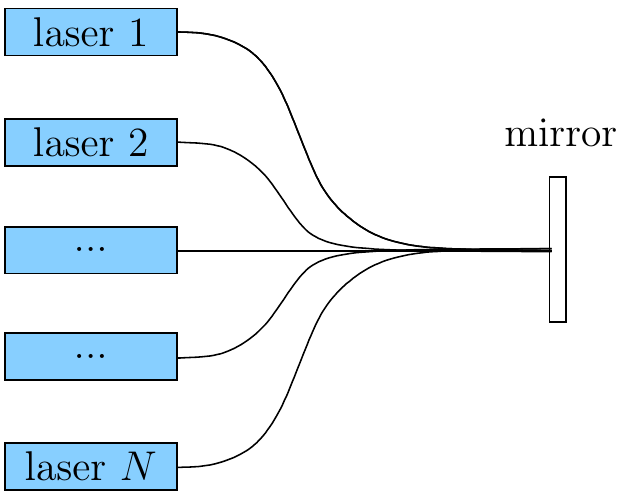}
  \caption{\label{fig:networkmirror}Coupling of multiple lasers via a
    common mirror.}
\end{SCfigure}

We will now discuss one promising all-optical setup that should, in
principle, be robust to phase mismatches. Consider the setup
\cite{AMA08} sketched in Fig.~\ref{fig:networkmirror}.  The laser
fields are coupled into a common fiber, which is terminated by a
mirror. As before, we assume that the coupling delays are equal (the
light paths are equally long on a cm scale), but allow for coupling
phases, i.e., differences in optical path-lengths on wavelength
scales.
In this setup, the connection from each laser~$k$ to the mirror
corresponds to a certain optical path length with a corresponding
phase $\psi_k$. The coupling phase from laser~$j$ to laser~$k$ is then
given by
\begin{equation}
  \phi_{kj} = \psi_k + \psi_j\;. \label{eq:phasesum}
\end{equation}
Similarly, the connection from each laser to the mirror has an
associated coupling strength $c_k$, which could in an experiment be
controlled by an attenuator in the corresponding fiber.  The coupling
from laser~$j$ to laser~$k$ then has an effective coupling strength
\begin{equation}
  \kappa_{kj} = c_k\cdot c_j \;. \label{eq:couplingsum}
\end{equation}

Under these conditions, the row sum is given by
\begin{equation}
  \sigma_k = \sum_{j=1}\kappa_{kj}\, e^{i(\phi_{kj}+u_k-u_j)} = c_k e^{i(\psi_k +u_k)} \sum_{j=1}^N c_j \, e^{i(\psi_j - u_j)} \;.
\end{equation}
The sum on the right hand side is a complex number independent of $k$.
The row sum condition can thus be satisfied if the prefactor is also
independent of $k$, i.e., all coupling strengths are equal ($c_k=c$) and
the relative phases $u_k$ are chosen by the system to give%
\footnote{The special role of laser 1 stems from our choice $u_1=0$.}
\begin{equation}
  \psi_1 = \psi_k + u_k \qquad(k=2,\dots,N)\;.
\end{equation}
In essence, all lasers couple to the optical mean field and compensate
for the difference in optical path length of their individual fiber by
adjusting their relative phase shift. As long as the setup obeys
Eqs.~(\ref{eq:phasesum}) and (\ref{eq:couplingsum}), the coupling
phases $\psi_k$ may vary, e.g., due to thermal effects, and do not
need to be controlled.

Assuming that all coupling strengths are equal and the relative phases
$u_k$ are tuned appropriately by the system the coupling matrix is
given by
\begin{equation}
  G_{kj} = c^2\, e^{i(\psi_k + \psi_j + u_k - u_j)} = c^2 \, e^{i2\psi_j} \;.
\end{equation}
This matrix has one eigenvalue corresponding to the row sum
\begin{equation}
  \sigma = c^2 \sum_{j=1}^N e^{i2\psi_j}
\end{equation}
and the $N-1$ transversal eigenvalues  are zero
\begin{equation}
  \gamma_{n} = 0 \qquad (n=1,\dots,N-1) \;,
\end{equation}
such that this setup is optimal for synchronization.

Recently, this coupling scheme has been proposed for optical
information processing using multi-mode lasers \cite{AMA08}. For
multi-mode lasers, the same argument as above holds for each
mode, such that the setup is, in principle, robust to phase
mismatches.  However, it may still be difficult to realize the
assumptions Eq.~(\ref{eq:phasesum}) and Eq.~(\ref{eq:couplingsum}) in
an experimental setup.

\section{Conclusion and outlook}
\label{sec:conclusions}
We have discussed chaos synchronization conditions for all-optically coupled
lasers. In all-optical coupling, the coupling phases play a crucial
role for the synchronizability.  The condition of constant row sum
corresponds to specific interference conditions, i.e., the input signals of
each laser should interfere in such a way that each laser receives the same input
signal, relative to its own phase. This corresponds to the existence
of an identical synchronization manifold.  Through interference, the
phases may compensate for mismatches in the coupling strengths.

Using these interference arguments we have explained experimental findings
\cite{PEI02} and discussed necessary and sufficient conditions for synchronization
of all network motifs which contain two all-optically coupled lasers.

Further, we have considered synchronization of larger laser networks,
and singled out the difficulties that arise in all-optical coupling
schemes due to the interference conditions. We predict that a setup of
all-to-all coupling via a common mirror may under certain conditions
be robust to phase mismatches and thus be optimal with respect to
stability of the synchronized chaotic dynamics.

In a broader context, these results might also be relevant for other networks
where phase-sensitive couplings play a role, e.g., networks of Stuart-Landau
oscillators with complex coupling constants $\sigma=Ke^{i\beta}$ \cite{CHO09}.
These are generic models representative for a large class of oscillator networks.

\section*{Acknowledgments}
We thank Ingo Fischer and Andreas Amann for helpful discussions.  This
work was supported by DFG in the framework of Sfb 910.  VF gratefully
acknowledges financial support from the German Academic Exchange
Service (DAAD).

\begin{acronym}[MSF]
  \acro{MSF}{master stability function}
\end{acronym}

\vfill

\section*{References}

\begin{thebibliography}{10}

\bibitem{PIK01}
Arkady~S. Pikovsky, Michael~G. Rosenblum, and J.~Kurths.
\newblock {\em Synchronization, A Universal Concept in Nonlinear Sciences}.
\newblock Cambridge University Press, Cambridge, 2001.

\bibitem{BOC02}
S.~Boccaletti, J.~Kurths, G.~Osipov, D.~L. Valladares, and C.~S. Zhou.
\newblock The synchronization of chaotic systems.
\newblock {\em Phys. Rep.}, 366:1--101, 2002.

\bibitem{CUO93}
K.~M. Cuomo and A.~V. Oppenheim.
\newblock {Circuit implementation of synchronized chaos with applications to
  communications}.
\newblock {\em Phys. Rev. Lett.}, 71(1):65--68, 1993.

\bibitem{KAN08a}
I.~Kanter, E.~Kopelowitz, and W.~Kinzel.
\newblock Public channel cryptography: chaos synchronization and {H}ilbert's
  tenth problem.
\newblock {\em Phys. Rev. Lett.}, 101(8):84102, 2008.

\bibitem{ARG05}
A.~Argyris, D.~Syvridis, L.~Larger, V.~Annovazzi-Lodi, P.~Colet, I.~Fischer,
  J.~Garc{\'i}a-Ojalvo, C.~R. Mirasso, L.~Pesquera, and K.~A. Shore.
\newblock {Chaos-based communications at high bit rates using commercial
  fibre-optic links}.
\newblock {\em Nature}, 438:343--346, 2005.

\bibitem{DHA04}
M.~Dhamala, V.~K. Jirsa, and M.~Ding.
\newblock Enhancement of neural synchrony by time delay.
\newblock {\em Phys. Rev. Lett.}, 92(7):074104, 2004.

\bibitem{CHO09}
C.~U. Choe, T.~Dahms, P.~H{\"o}vel, and E.~Sch{\"o}ll.
\newblock Controlling synchrony by delay coupling in networks: from in-phase to
  splay and cluster states.
\newblock {\em Phys. Rev.~E}, 81(2):025205(R), 2010.

\bibitem{KIN09}
W.~Kinzel, A.~Englert, G.~Reents, M.~Zigzag, and I.~Kanter.
\newblock Synchronization of networks of chaotic units with time-delayed
  couplings.
\newblock {\em Phys. Rev.~E}, 79(5):056207, 2009.

\bibitem{KAN11}
I.~Kanter, M.~Zigzag, A.~Englert, F.~Geissler, and W.~Kinzel.
\newblock Synchronization of unidirectional time delay chaotic networks and the
  greatest common divisor.
\newblock {\em Europhys.~Lett.}, 93(6):60003, 2011.

\bibitem{WUE05a}
H.~J. W{\"u}nsche, S.~Bauer, J.~Kreissl, O.~Ushakov, N.~Korneyev,
  F.~Henneberger, E.~Wille, H.~Erzgr{\"a}ber, M.~Peil, W.~Els{\"a}{\ss}er, and
  I.~Fischer.
\newblock Synchronization of delay-coupled oscillators: A study of
  semiconductor lasers.
\newblock {\em Phys.~Rev.~Lett.}, 94:163901, 2005.

\bibitem{CAR06}
T.~W. Carr, I.~B. Schwartz, M.~Y. Kim, and R.~Roy.
\newblock Delayed-mutual coupling dynamics of lasers: scaling laws and
  resonances.
\newblock {\em SIAM J.~Appl. Dyn. Syst.}, 5(4):699--725, 2006.

\bibitem{ERZ06a}
H.~Erzgr{\"a}ber, B.~Krauskopf, and D.~Lenstra.
\newblock Compound laser modes of mutually delay-coupled lasers.
\newblock {\em SIAM J.~Appl. Dyn. Syst.}, 5(1):30--65, 2006.

\bibitem{FIS06}
I.~Fischer, R.~Vicente, J.~M. Buld{\'u}, M.~Peil, C.~R. Mirasso, M.~C. Torrent,
  and J.~Garc{\'i}a-Ojalvo.
\newblock Zero-lag long-range synchronization via dynamical relaying.
\newblock {\em Phys.~Rev.~Lett.}, 97(12):123902, 2006.

\bibitem{DHU08}
O.~{D'Huys}, R.~Vicente, T.~Erneux, J.~Danckaert, and I.~Fischer.
\newblock Synchronization properties of network motifs: Influence of coupling
  delay and symmetry.
\newblock {\em Chaos}, 18(3):037116, 2008.

\bibitem{FLU09}
V.~Flunkert, O.~{D'Huys}, J.~Danckaert, I.~Fischer, and E.~Sch{\"o}ll.
\newblock Bubbling in delay-coupled lasers.
\newblock {\em Phys. Rev.~E}, 79:065201 (R), 2009.

\bibitem{ZAM10}
Jordi Zamora-Munt, C.~Masoller, Jordi Garcia-Ojalvo, and R.~Roy.
\newblock Crowd synchrony and quorum sensing in delay-coupled lasers.
\newblock {\em Phys. Rev. Lett.}, 105(26):264101, 2010.

\bibitem{HIC11}
K.~Hicke, O.~{D'Huys}, V.~Flunkert, E.~Sch{\"o}ll, J.~Danckaert, and
  I.~Fischer.
\newblock Mismatch and synchronization: Influence of asymmetries in systems of
  two delay-coupled lasers.
\newblock {\em Phys. Rev.~E}, 83:056211, 2011.

\bibitem{AVI12}
Y.~Aviad, I.~Reidler, M.~Zigzag, M.~Rosenbluh, and I.~Kanter.
\newblock Synchronization in small networks of time-delay coupled chaotic diode
  lasers.
\newblock {\em Opt. Express}, 20(4):4352--4359, 2012.

\bibitem{AVI08}
Y.~Aviad, I.~Reidler, W.~Kinzel, I.~Kanter, and M.~Rosenbluh.
\newblock Phase synchronization in mutually coupled chaotic diode lasers.
\newblock {\em Phys.~Rev.~E}, 78(2):025204, 2008.

\bibitem{PEI02}
M.~Peil, T.~Heil, I.~Fischer, and W.~Els{\"a}{\ss}er.
\newblock Synchronization of chaotic semiconductor laser systems: A vectorial
  coupling-dependent scenario.
\newblock {\em Phys. Rev. Lett.}, 88(17):174101, 2002.

\bibitem{PEC98}
L.~M. Pecora and T.~L. Carroll.
\newblock Master stability functions for synchronized coupled systems.
\newblock {\em Phys. Rev. Lett.}, 80(10):2109--2112, 1998.

\bibitem{FLU10b}
V.~Flunkert, S.~Yanchuk, T.~Dahms, and E.~Sch{\"o}ll.
\newblock Synchronizing distant nodes: a universal classification of networks.
\newblock {\em Phys.~Rev.~Lett.}, 105:254101, 2010.

\bibitem{HEI11}
S.~Heiligenthal, T.~Dahms, S.~Yanchuk, T.~J{\"u}ngling, V.~Flunkert, I.~Kanter,
  E.~Sch{\"o}ll, and W.~Kinzel.
\newblock Strong and weak chaos in nonlinear networks with time-delayed
  couplings.
\newblock {\em Phys. Rev. Lett.}, 107:234102, 2011.

\bibitem{SUN09a}
J.~Sun, E.~M. Bollt, and T.~Nishikawa.
\newblock Master stability functions for coupled nearly identical dynamical
  systems.
\newblock {\em Europhys. Lett.}, 85(6):60011, 2009.

\bibitem{SOR11b}
F.~Sorrentino and M.~Porfiri.
\newblock Analysis of parameter mismatches in the master stability function for
  network synchronization.
\newblock {\em EPL}, 93(5):50002, 2011.

\bibitem{KOC00a}
L.~Kocarev, Ulrich Parlitz, and Reggie Brown.
\newblock {Robust synchronization of chaotic systems}.
\newblock {\em Phys. Rev. E}, (4):3716--3720, 2000.

\bibitem{FAR82}
J.~D. Farmer.
\newblock Chaotic attractors of an infinite-dimensional dynamical system.
\newblock {\em Physica~D}, 4:366, 1982.

\bibitem{GIA96}
G.~Giacomelli and A.~Politi.
\newblock Relationship between delayed and spatially extended dynamical
  systems.
\newblock {\em Phys.~Rev.~Lett.}, 76:2686, 1996.

\bibitem{MEN98b}
B.~Mensour and Andr\'e Longtin.
\newblock Power spectra and dynamical invariants for delay-differential and
  difference equations.
\newblock {\em Physica~D}, 113:1--25, 1998.

\bibitem{WOL06}
M.~Wolfrum and S.~Yanchuk.
\newblock Eckhaus instability in systems with large delay.
\newblock {\em Phys.~Rev.~Lett.}, 96:220201, 2006.

\bibitem{YAN06}
S.~Yanchuk, M.~Wolfrum, P.~H{\"o}vel, and E.~Sch{\"o}ll.
\newblock Control of unstable steady states by long delay feedback.
\newblock {\em Phys.~Rev.~E}, 74:026201, 2006.

\bibitem{YAN09}
S.~Yanchuk and P.~Perlikowski.
\newblock Delay and periodicity.
\newblock {\em Phys. Rev. E}, 79(4):046221, 2009.

\bibitem{ILL11}
L.~Illing, Cristian~D. Panda, and Lauren Shareshian.
\newblock Isochronal chaos synchronization of delay-coupled optoelectronic
  oscillators.
\newblock {\em Phys. Rev. E}, 84:016213, 2011.

\bibitem{FLU11a}
V.~Flunkert.
\newblock {\em Delay-Coupled Complex Systems}.
\newblock Springer Theses. Springer, Heidelberg, 2011.

\bibitem{VIC02}
R.~Vicente, T.~P\'erez, and C.~R. Mirasso.
\newblock {Open-versus closed-loop performance of synchronized chaotic
  external-cavity semiconductor lasers}.
\newblock {\em IEEE J.~Quantum Electron.}, 38(9):1197--1204, 2002.

\bibitem{VOS00}
Henning~U. Voss.
\newblock Anticipating chaotic synchronization.
\newblock {\em Phys. Rev. E}, 61(5):5115--5119, 2000.

\bibitem{MAS01}
C.~Masoller.
\newblock Anticipation in the synchronization of chaotic semiconductor lasers
  with optical feedback.
\newblock {\em Phys. Rev. Lett.}, 86(13):2782--2785, 2001.

\bibitem{KOZ00}
G.~Kozyreff, A.~G. Vladimirov, and P.~Mandel.
\newblock Global coupling with time delay in an array of semiconductor lasers.
\newblock {\em Phys.~Rev.~Lett.}, 85(18):3809, 2000.

\bibitem{WIE09a}
S.~Wieczorek.
\newblock Stochastic bifurcation in noise-driven lasers and hopf oscillators.
\newblock {\em Phys. Rev. E}, 79(3):036209, 2009.

\bibitem{AMA08}
A.~Amann, A.~Pokrovskiy, S.~Osborne, and S.~O'Brien.
\newblock Complex networks based on discrete-mode lasers.
\newblock In {\em International Workshop on Multi-Rate Processes and
  Hysteresis, Cork 2008}, volume 138 of {\em J. Phys.: Conf. Ser.}, page
  012001. IOP Publishing, 2008.

\bibitem{APP11}
L.~Appeltant, M.~C. Soriano, Guy Van~der Sande, J.~Danckaert, S.~Massar,
  J.~Dambre, B.~Schrauwen, C.~R. Mirasso, and I.~Fischer.
\newblock Information processing using a single dynamical node as complex
  system.
\newblock {\em Nat. Commun.}, 2:468, 2011.

\bibitem{SCH06a}
S.~Schikora, P.~H{\"o}vel, H.~J. W{\"u}nsche, E.~Sch{\"o}ll, and
  F.~Henneberger.
\newblock All-optical noninvasive control of unstable steady states in a
  semiconductor laser.
\newblock {\em Phys.~Rev.~Lett.}, 97:213902, 2006.

\bibitem{UDA01}
Vladimir~S. Udaltsov, J.~P. Goedgebuer, L.~Larger, and William~T. Rhodes.
\newblock Communicating with optical hyperchaos: Information encryption and
  decryption in delayed nonlinear feedback systems.
\newblock {\em Phys. Rev. Lett.}, 86(9):1892--1895, 2001.

\bibitem{CAL09}
K.~E. Callan, L.~Illing, Z.~Gao, D.~J. Gauthier, and E.~Sch{\"o}ll.
\newblock Broadband chaos generated by an opto-electronic oscillator.
\newblock {\em Phys.~Rev.~Lett.}, 104(11):113901, 2010.

\bibitem{RAV11}
Bhargava Ravoori, Adam~B. Cohen, Jie Sun, Adilson~E. Motter, Thomas~E. Murphy,
  and R.~Roy.
\newblock Robustness of optimal synchronization in real networks.
\newblock {\em Phys. Rev. Lett.}, 107:034102, 2011.

\bibitem{ROG01b}
F.~Rogister, A.~Locquet, D.~Pieroux, M.~Sciamanna, O.~Deparis, P.~Megret, and
  M.~Blondel.
\newblock {Secure communication scheme using chaotic laser diodes subject to
  incoherent optical feedback and incoherent optical injection}.
\newblock {\em Opt. Lett.}, 26(19):1486--1488, 2001.

\bibitem{SUK04}
David~W. Sukow, Karen~L. Blackburn, Allison~R. Spain, Katherine~J. Babcock,
  Jake~V. Bennett, and Athanasios Gavrielides.
\newblock {Experimental synchronization of chaos in diode lasers with
  polarization-rotated feedback and injection}.
\newblock {\em Opt. Lett.}, 29(20):2393, 2004.

\end{thebibliography}

\end{document}